\documentclass[twoside]{article}
\usepackage{fleqn,espcrc2,epsfig}

\title{{
\begin{flushright}
\vskip -20pt
{\normalsize
PDK-752 \\
TUHEP-00-05 \\
\vskip -6pt
17 July 2000}
\end{flushright}
{
New results on atmospheric neutrinos from Soudan 2}}
\thanks{Presented at NU2000, the XIXth Int. Conference on Neutrino Physics
and Astrophysics, June 16 - 21, 2000, Sudbury, Canada.}}

\author{W. Anthony Mann \thanks{High Energy Physics, Tufts University,
4 Colby St., Medford, MA 02155, USA}
\vskip 0.2cm
\noindent {\it for the Soudan-2 Collaboration}
\vskip 0.1cm
\noindent (Argonne National Laboratory, University of Minnesota, Tufts
University, Western Washington, USA; Oxford University, Rutherford Appleton 
Laboratory, UK)}

\newcommand {\numu}{$\nu_{\mu}$}
\newcommand {\nue}{$\nu_{e}$}
\newcommand {\nutau}{$\nu_{\tau}$}

\newcommand {\delm}{$\Delta m^{2}$}
\newcommand {\sqsin}{$\sin^{2} 2\theta$}
\newcommand {\loe}{$L/E$}
\newcommand {\loenu}{$L/E_{\nu}$}

\newcommand {\sqchi}{$\chi^{2}$}

\newcommand {\nmflvr}{\mbox{$\nu_\mu$\ flavor}}

\newcommand {\neflvr}{\mbox{$\nu_e$-flavor}}
\newcommand {\dlmval}[1]{\ensuremath{\mbox{$\Delta m^{2} = #1$} \ 
\mbox{eV}^{2}}}

\begin{document}

\begin{abstract} Neutrino interactions recorded in a 5.1 fiducial kiloton-year 
exposure of the Soudan-2 iron tracking calorimeter are analyzed for effects of
neutrino oscillations.   Using contained single track and single shower events, 
we update our measurement of the atmospheric \numu /\nue\ ratio-of-ratios and 
find $R = 0.68 \pm 0.11 \pm 0.06$.   Assuming this anomalously low R-value is 
the result of \numu\ flavor disappearance via \numu\ to \nutau\ oscillation, we 
select samples of charged current events which offer good resolution, 
event-by-event, for \loenu\ reconstruction.  Oscillation-weighted Monte Carlo 
events are fitted to these data events using a $\chi^2$ function summed over 
bins of log$(L/E_{\nu})$.  The region allowed 
in the (\sqsin , \delm ) plane at 90\% CL is obtained 
using the Feldman-Cousins procedure: 
$0.46 < \sin^{2} 2\theta \leq 1.0$ and $2.2 \times 10^{-4} < \Delta m^{2} < 
2.2 \times 10^{-2} \mbox{eV}^{2}$.  A small but relatively energetic sample of
partially contained \numu\ events has also been isolated.  Their distribution in
log$(L/E_{vis})$ relative to null oscillation Monte Carlo is  
compatible with \numu\ to \nutau\ oscillation scenarios within the parameters 
region allowed by our contained events.
\end{abstract}

\maketitle

\section{DETECTOR; DATA EXPOSURE}


    The Soudan-2 experiment is currently taking data using its fine-grained
iron tracking cal\-or\-imeter of total mass 963 tons.  This detector images
non-relativistic as well as relativistic charged particles produced in 
atmospheric neutrino reactions.  It is operating underground 
at a depth of 2100 meters-water-equivalent on level 27 of the Soudan 
Mine State Park in northern Minnesota (northwest of Sudbury).  
The calorimeter's modular design 
enabled data-taking to commence in April 1989 when the detector was one 
quarter of its full size; assembly of the detector was completed during 1993.  
Data-taking has continued with 85\% live time, even though dynamite 
blasting has been underway nearby for the MINOS cavern excavation since
Summer 1999.  The data exposure as of this Conference is 5.40 fiducial 
kiloton-years (kty). Results presented here are based upon a 5.1 kty exposure.

      The tracking calorimeter operates as a slow-drift (0.6 cm/$\mu$s)
time projection chamber.  Its tracking elements are meter-long plastic
drift tubes which are placed into the corrugations of steel sheets.  The
sheets are stacked to form a tracking lattice of honeycomb geometry.  A
stack is packaged as a calorimeter module and the detector is assembled
building-block fashion using these modules~\cite{NIM}. The calorimeter is 
surrounded on all sides by a cavern-liner active shield array of two or three 
layers of proportional tubes ~\cite{NIM_VS}.

      Topologies for contained events in Soudan 2 include single track and 
single shower events (mostly \numu\ and \nue\ quasi-elastic reactions) and 
multiprong events.  Flavor-tagging proceeds straightforwardly:  An
event having a leading, non-scattering track with ionization $dE/dx$
compatible with muon mass is a candidate charged current (CC) event of
\nmflvr ; an event having a prompt, relatively energetic shower prong
is a candidate $\nu_{e}$ CC event.
Recoil protons of momenta greater than approx. 350 MeV/$c$ are imaged by the 
calorimeter, allowing a much-improved measurement of the incident neutrino 
direction, especially for sub-GeV quasi-elastic reactions.

\section{ATMOSPHERIC {\bf $\nu$} FLAVOR RATIO}

      We measure the atmospheric neutrino  \numu /\nue\ flavor ratio-of-ratios
$R$ using single track and single shower events which are fully contained within
the calorimeter (all hits more than 20 cm from the nearest surface). 
These samples contain mostly quasi-elastic neutrino reactions, but include a 
background of photon and neutron reactions originating in cosmic ray muon 
interactions in the surrounding cavern rock.  The latter ``rock events" are 
mostly tagged by coincident hits in the active shield, however some are 
unaccompanied by shield hits and constitute a background. The amount of 
zero-shield-hit rock background in a neutrino event sample is estimated by 
fitting event vertex-depth distributions to a combination of tagged-rock and 
$\nu$ Monte Carlo distributions.  As expected, the fits show the background to 
be mostly confined to outer regions of the calorimeter.  Details of our 
analysis procedures for quasi-elastic events can be found in 
Refs.~\cite{Allison2}.

      The track and shower event samples for our 5.1 kty exposure are summarized
in Table~\ref{tab:results}. Our full detector Monte Carlo simulation of
atmospheric neutrino interactions is based
on the 1996 Bartol flux for the Soudan site~\cite{VAgrawal}. 

\begin{table}[hbt]
\caption{Soudan-2 track and shower event samples from the 5.1 fiducial 
kiloton-year exposure.}
\setlength{\tabcolsep}{0.7pc}
\begin{tabular}{@{}l@{\extracolsep{\fill}}cc}
\hline
 & Tracks & Showers \\
\hline
Data, raw & 133 & 193 \\
Monte Carlo events & 1097 & 1017 \\
(norm. to 5.1 kty) & 193.1 & 179.0 \\
Data, bkgrd subtr. & 105.1$\pm$12.7 & 142.3$\pm$13.9 \\
\hline
\end{tabular}
\label{tab:results}
\end{table}

After correction for cosmic ray muon induced background, the number 
of single track events observed in data is less than the number of single 
shower events, whereas the null oscillation Monte Carlo predicts the relative 
rates to be other-way-round. Consequently the flavor ratio-of-ratios 
obtained is less than 1.0 and is anomalous: 
$$ R = 0.68 \pm 0.11(stat) \pm 0.06 (sys).$$

This value is equal to the $R$ value obtained
last summer using 4.6 kty exposure~\cite{plenary}.
         
\section{SAMPLE FOR {\bf \loe } MEASUREMENT}

The phenomenology for \numu\ to \nutau\ oscillations is quite specific;
neutrinos of muon flavor can metamorphose and 
thereby ``disappear" according to the equation
\begin{eqnarray}
\lefteqn{P(\nu_{\mu} \rightarrow \nu_{\tau}) = } \nonumber \\
& & \sin^2 (2\theta) \cdot \sin^2 \left[ \frac{1.27 \mbox{ } \Delta m^{2} 
[\mbox{eV}^2] \cdot L \mbox{[km] }}{E_\nu \mbox{[GeV] }} \right]
\end{eqnarray}
Consequently it is optimal to analyze for neutrino oscillations using the
variable \loenu .   With the Soudan-2 calorimeter, measurement of event
energy for charged current reactions is straightforward; we do this
with resolution $\Delta E/E$ which is 20\% for \numu\ CC's and 23\% for 
\nue\ CC's.  To determine the neutrino path length $L$ however, 
the zenith angle
$\theta_{z}$ of the incident neutrino must be reconstructed with accuracy. 
The path length can be calculated from the zenith angle according to
\begin{eqnarray}
\lefteqn{L(\theta_z) = } \nonumber \\
& & \sqrt{(R - d)^2 \cos^2\theta_z + (d + h)(2R - d + h)} \nonumber \\
& & - (R - d)\cos\theta_z
\end{eqnarray}
where $R$ is the Earth's radius, $d$ is the depth of the detector, and $h$ is
the mean neutrino production height.  The
latter is a function of $\nu$ flavor, $\nu$ energy, and
$\theta_z$~\cite{Ruddick}.

We select from our data an event sample suited to this measurement.  We use 
a quasi-elastic track or shower event provided that the recoil proton is 
measured and that $P_{lept}$ exceeds 150 MeV/$c$; otherwise, if the recoil 
nucleon is not visible, we require the single lepton to have $E_{vis}$ great 
than 600 MeV. We also select multiprong events, provided they are energetic 
($E_{vis}$ greater than 700 MeV) and have vector sum of $P_{vis}$ exceeding
450 MeV/$c$ (to ensure clear directionality). Additionally, the final state
lepton momenta are required to exceed 250 MeV/$c$.
For the selected sample, flavor tagging is 
estimated to be correct for more than 92\% of events.  The resolution for 
recovering the incident neutrino direction is evaluated using the mean angular
separation between ``true" versus reconstructed neutrino direction in
Monte Carlo events.  The mean separations are $33.2^{\circ}$ for \numu\ CC's and
$21.3^{\circ}$ for \nue\ CC's .  The resolution 
in $\log$ \loenu\ ($L$ in kilometers,
$E_\nu$ in GeV) is better than 0.5 for the selected sample. Hereafter we
refer to these events as ``HiRes events".

The zero-shield-hit rock background, as estimated by the fits to event
vertex depth distributions, comprises 6.8\% (5.1\%) of the \numu\ (\nue )
flavor sample of HiRes events.

Table~\ref{tab:HiRes} shows the HiRes event populations.  After
background subtraction there are 106.3 data events of \nmflvr\ and 132.8 
events of \neflvr . Using these events, whose mean energy is higher than 
that of our track and shower events, the ratio-of-ratios 
is $R = 0.67 \pm 0.12$, which is also significantly less than 1.0.

\begin{table}[hbt]
\caption{Event samples selected for good \loenu\ resolution, including 
atmospheric $\nu$ data (without, with background subtraction) and $\nu$ 
Monte Carlo samples. The MC rates are shown normalized to the 
\nue\ CC data.}
\setlength{\tabcolsep}{0.7pc}
\begin{tabular}{@{}l@{\extracolsep{\fill}}cccc}
\hline
& & & \numu\ & \nue\ \\
\hline
 Data, raw & & & 114.0$\pm$10.7 & 140.0$\pm$11.8 \\
 Data, subt. & & & 106.3$\pm$14.7 & 132.8$\pm$13.4 \\
 Monte Carlo & & & 158.5$\pm$ 4.8 & 132.8$\pm$4.4 \\
\hline
\end{tabular}
\label{tab:HiRes}
\end{table}

\begin{figure}[htb]
\centerline{\epsfig{file=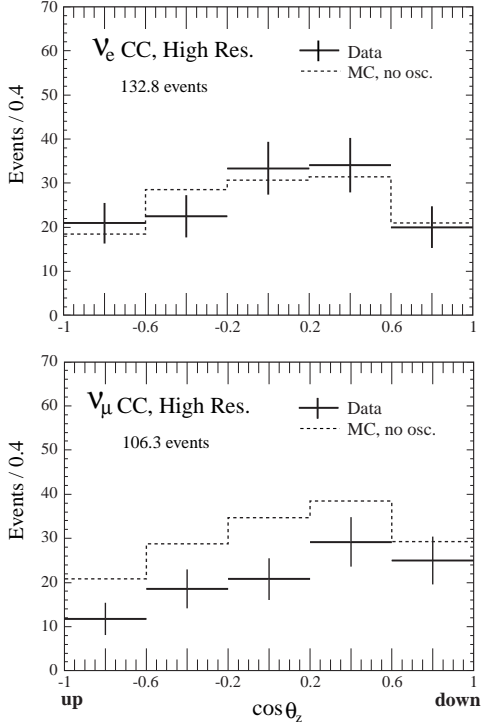,width=2.5in}}
\caption{Distributions of $\cos \theta_{z}$ for \nue\ and \numu\ flavor 
HiRes samples.  Data (crosses) are compared to the null oscillation Monte 
Carlo (dashed histograms) where the MC has been rate-normalized to the \nue\ 
data.}
\label{fig:fig01}
\end{figure}

\begin{figure}[htb]
\centerline{\epsfig{file=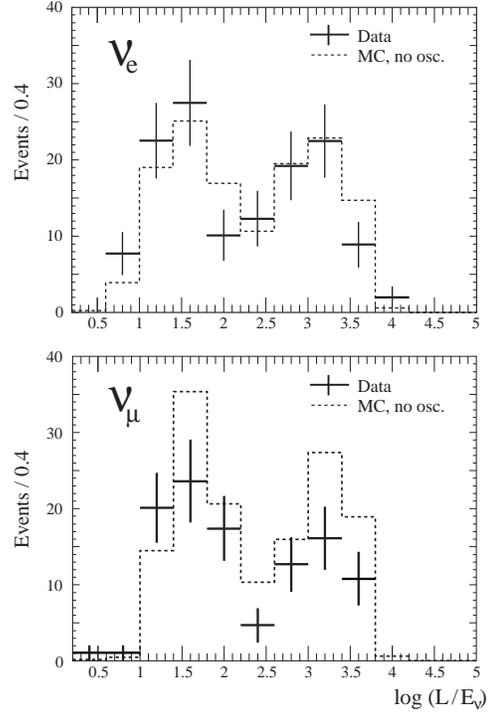,width=2.5in}}
\caption{Distributions of $\log\left(L/E_{\nu}\right)$ 
for \nue\ and \numu\ charged
current events compared to the atmospheric neutrino MC with no oscillations.
The MC is shown rate-normalized to the \nue\ data.}
\label{fig:fig02}
\end{figure}

The atmospheric Monte Carlo (MC) sample
represents 28.2 kiloton years of exposure.   The MC event rates displayed 
in Table~\ref{tab:HiRes} have been normalized to the \nue\ data sample.   This
normalization is equivalent to a reduction of the Bartol neutrino fluxes by
21\%.  The assumption implicit with this adjustment is that the \nue\ 
sample is devoid of oscillation effects.
Figs.~\ref{fig:fig01},~\ref{fig:fig02}~and~\ref{fig:fig03} show HiRes
distributions with this normalization in place.

Fig.~\ref{fig:fig01} shows the distributions of these samples in
cosine of the zenith angle.   For \nue\ events, the shape of the 
distribution for data (Fig.~\ref{fig:fig01}a, crosses) coincides with that
predicted by the Monte Carlo (dashed histogram) for null oscillations.
The distribution of \numu\ data however, falls below the MC prediction in 
all bins (Fig.~\ref{fig:fig01}b) with the relative dearth being more pronounced
for \numu 's incident from below horizon.   
Distributions in $\log\left(L/E_{\nu}\right)$
for HiRes events are shown in Fig.~\ref{fig:fig02}.   For null oscillations
this variable distributes according to a `phase space' which reflects the
neutrino points-of-origin throughout the spherical shell volume of the Earth's
atmosphere.  That is, down-going $\nu$'s populate the peak at lower
$\log\left(L/E_{\nu}\right)$ from 0.0 to 2.0.  Neutrinos incident from/near
the horizon occur within the dip region extending from 2.0 to 2.6, while
upward-going neutrinos populate the peak at higher values.
Fig.~\ref{fig:fig02}a shows that, allowing for statistical fluctuations,
the \nue\ data follows the shape of the null oscillation MC distribution.
In contrast, the \numu\ data (Fig.~\ref{fig:fig02}b) falls below the null
ocillation MC for all but the most vertically down-going flux.

\section{{\bf ($\sin^{2} 2\theta$, $\Delta m^{2}$)} ALLOWED REGION}

\begin{figure}[htb]
\centerline{\epsfig{file=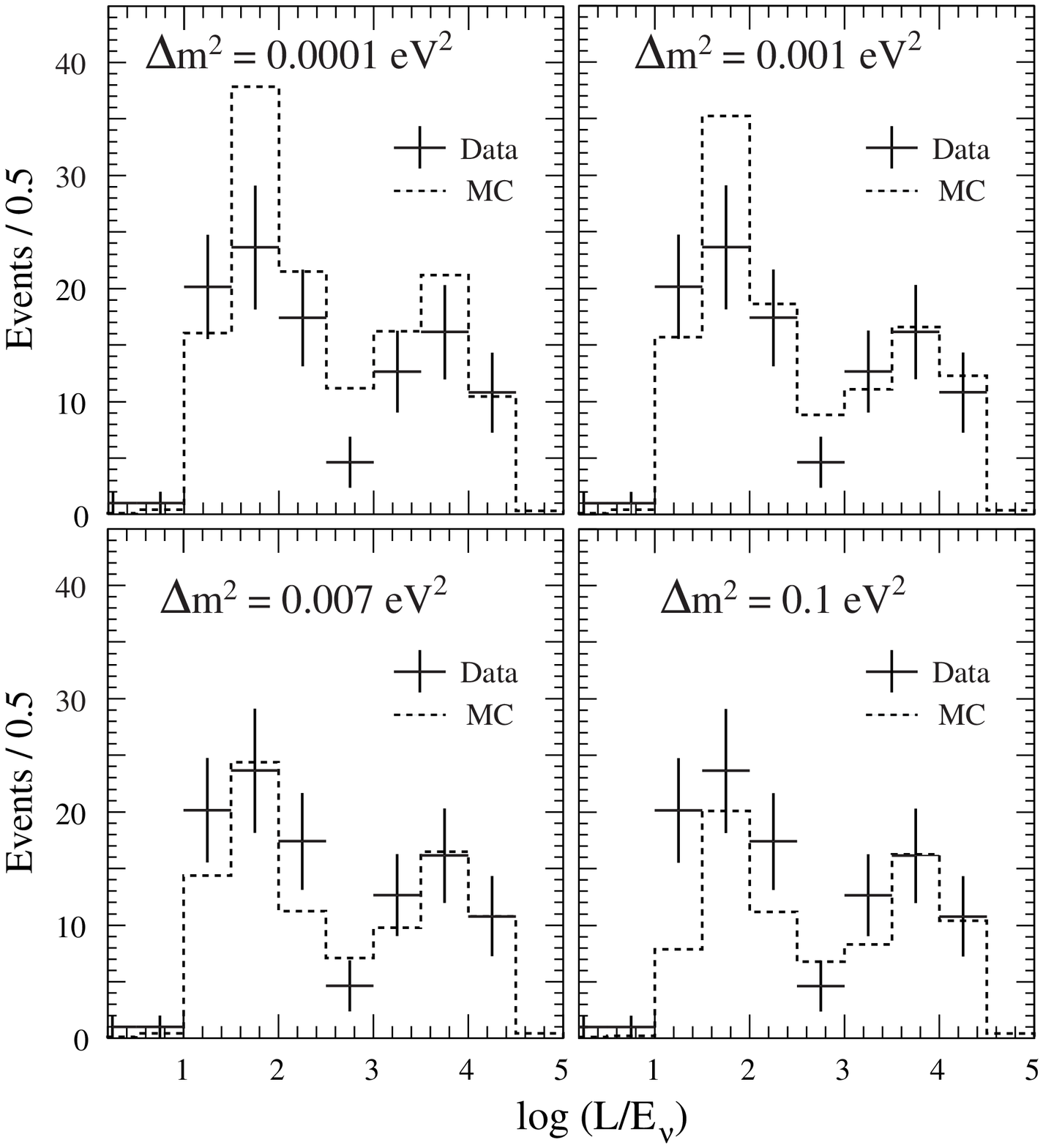,width=2.75in}}
\caption{Comparison of $\log\left(L/E_{\nu}\right)$ distribution for \numu\ data
(crosses) with expectations for \mbox{\numu  $\leftrightarrow$ \nutau }
oscillation with \sqsin\ = 1 (dashed histograms), for four different \delm\ 
values.}
\label{fig:fig03}
\end{figure}  

To convert results of our atmospheric neutrino simulation generated under the
no-oscillation hypothesis into simulated neutrino oscillation data, we apply to
every MC event an \loenu -dependent weight representing the probability of
\nmflvr\ survival for a given \delm\ and \sqsin .

An exploratory matchup of \numu\ data with tri\-al oscillation scenarios is
shown in Fig.~\ref{fig:fig03}. For the mixing angle \sqsin\ fixed at unity, we
plot the MC distribution weighted for \numu\ to \nutau\ oscillations with 
differing \delm\ settings.  With \dlmval{10^{-4}} (Fig.~\ref{fig:fig03}a), the 
oscillation prediction lies above the data in most $L/E_{\nu}$ bins.  With 
\dlmval{10^{-3}} (Fig.~\ref{fig:fig03}b), the upgoing $\nu$ flux is now better 
described by the MC, although the expectation for horizontal and down-going 
neutrinos remains somewhat high. With \dlmval{7 \times 10^{-3}}, 
a rough agreement overall is achieved (Fig.~\ref{fig:fig03}c). 
Going to higher \delm , we find that for \dlmval{10^{-1}} the 
oscillation-weighted MC falls below 
the data in most bins (Fig.~\ref{fig:fig03}d).  These hints concerning the 
parameters regime preferred by the data are borne out by a more 
considered analysis, as we now show.

To determine the neutrino oscillation parameters \delm\ and \sqsin\ from our 
data, we construct a \sqchi\ function over the plane-of-parameters. For points
$(i,j)$ in the physical region of the 
$(\sin^{2} 2\theta_{i}, \: \log \Delta m^{2}_{j})$ plane, 
we fit the MC expection to our data at each point. The MC flux 
normalization, $f_{\nu}$ as well as $\sin^{2} 2\theta_{i}$ and 
$\log \Delta m^{2}_{j}$, is a free parameter: 
\begin{eqnarray}
\lefteqn{(\chi^{2}_{data})_{ij} =
\ \chi^{2} (\sin^2 2\theta_i,\:\Delta m^2_j,\:f_{\nu})} \nonumber \\
\lefteqn{=\ 
\sum_{k=1}^{8}\frac{\left( N_{k}(data - bkgd) - f_{\nu} \cdot N_{k}(MC)
\right)^2}{\sigma^{2}_{k}}.}
\end{eqnarray}
We assume that the oscillation affecting our data is purely \numu\ into
\nutau\ and that the \nue\ data is unaffected.

The \sqchi\ is summed over data bins containing our selected (HiRes)
\numu\ and \nue\ samples, where $k = 1-7$ are \numu\ 
$\log\left(L/E_{\nu}\right)$ 
bins, with $k = 8$ containing all the \nue\ events.  The denominator
$\sigma_{k}^{2}$ accounts for finite statistics in the neutrino Monte Carlo 
and for uncertainty in the rock background in the $\nu$ data.  
Not yet included are error terms which address systematic errors 
in the analysis, however preliminary examination shows statistical
errors to be the dominant error source in the analysis. The MC counts $N_k(MC)$ 
for the $k^{th}$ bin are constructed using oscillation weight factors.
 
\begin{figure}[htb]
\centerline{\epsfig{file=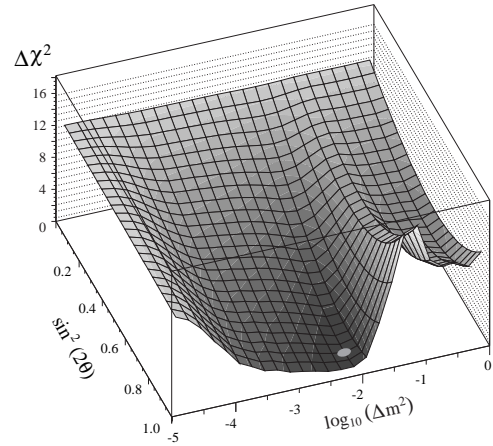,width=2.5in}}
\caption{The surface of $\Delta \chi^{2}$ over the \delm , \sqsin\ plane; the 
MC normalization is allowed to adjust at each point.  The oval at the bottom 
of the basin at large mixing angle denotes the $\chi^{2}_{min}$ location.}
\label{fig:fig04}
\end{figure}

We find the location of minimum $\chi^{2}_{data}$, and plot  
$( \Delta \chi^{2}_{data})_{ij}$ which is $( \chi^{2}_{data})_{ij} -
( \chi^{2}_{data})_{min}$.  The $\Delta \chi^{2}$ surface thereby obtained is
shown in Fig.~\ref{fig:fig04}.  A crater region of low $\chi^{2}$ 
values is clearly discerned, at the bottom of which is a relatively flat basin. 
The lowest point $\chi^{2}_{min}$ occurs at values \sqsin\ = 0.90,
\dlmval{7.9 \times 10^{-3}}, with flux normalization $f_{\nu}$ = 0.78.

An additional structure is the $\Delta \chi^{2}$ ridge which occurs at large
mixing angle and for \delm\ above 10$^{-2}$ eV$^{2}$ .  
For oscillation solutions 
in this regime, depletion in the downward-going \numu\ neutrino flux with
sub-GeV energies is predicted for \numu\ to \nutau\ oscillations by equation 
(1) arising from the first oscillation minimum.  Our HiRes events have 
sufficient resolution to show such an effect if it would be present. However,
no pronounced depletion is observed, and so the \sqchi\ has a high value 
there.

\begin{figure}[htb]
\centerline{\epsfig{file=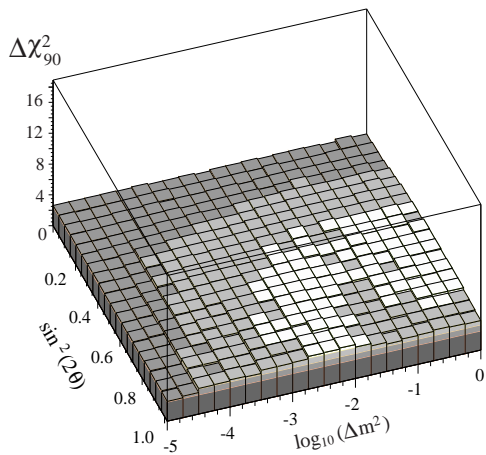,width=2.5in}}
\caption{The surface of $\Delta \chi_{90}^{2}$ over the parameters plane 
described via the Feldman-Cousins procedure.  The intersection of this 
surface with the $\Delta \chi^{2}$ surface of Fig.~\ref{fig:fig04} defines 
the parameters region allowed by Soudan-2 data at 90\% CL.}
\label{fig:fig05}
\end{figure}

    To find the region allowed for the oscillation parameters by our data
at 90\% confidence level (CL), we use the method of Feldman and
Cousins~\cite{Feldman}.  At each of 2500 points $(i,j) = (\sin^{2} 2\theta_{i}, 
\Delta m^{2}_{j})$ on a grid spanning the physical region of the plane
parameters, we run 1000 simulated experiments. For each of the simulated sets,
we find $(\Delta \chi_{90}^{2})_{ij}$ such that $(\Delta \chi_{sim}^{2})_{ij}$
is less than $(\Delta \chi_{90}^{2})_{ij}$ for 90\% of the simulated experiments
at $(i,j)$.  The surface defined by local $\Delta \chi_{90}^{2}$ over the 
oscillation parameters plane is shown in Fig.~\ref{fig:fig05}.  Note that the 
surface is not a plane at $\Delta \chi_{90}^{2}$ = 4.61, but rather has a
concave shape.  The central shaded portion is approximately
$\Delta \chi^{2}$ = 4.6, however the outlying regions have 
$\Delta \chi^{2}$ values which are lower. At each point over the
physical region, if $(\Delta \chi^{2}_{data})_{ij}$ is less than 
$(\Delta \chi^{2}_{90})_{ij}$, then $(i,j)$ belongs to the allowed region of 
the 90\% CL contour.

\begin{figure}[htb]
\centerline{\epsfig{file=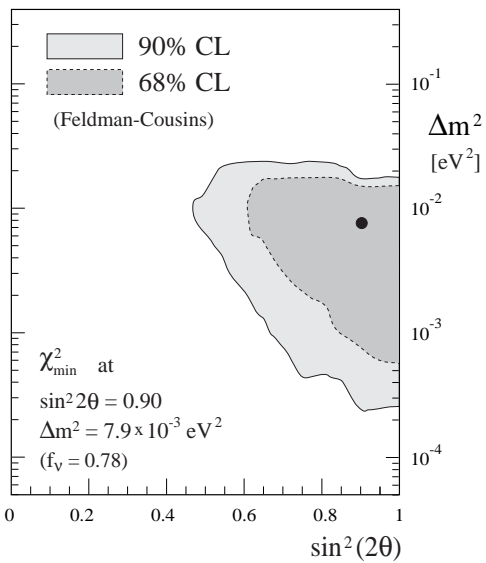,width=2.5in}}
\caption{Soudan-2 allowed regions at 68\% and 90\% CL with the best fit point 
for \numu\ $\rightarrow$ \nutau\ oscillations.}
\label{fig:fig06}
\end{figure}

   The region allowed by our data at 90\% CL is shown by the shaded area in 
Fig.~\ref{fig:fig06}.  Although $\chi^{2}_{min}$ occurs at the location 
depicted by the solid circle, the relatively flat basin of our $\Delta \chi^{2}$
surface extends to lower \delm\ values. SuperK has reported their best fit 
\delm\ value to be \mbox{$3.2 \times 10^{-3}~\mbox{eV}^{2}$}~\cite{Sobel}; our
data is compatible with that as well as with somewhat higher \delm\ values.

\section{PARTIALLY CONTAINED EVENTS}

We plan to include more data in the above analysis.
An additional data sample is comprised of \numu\ flavor events which
are partially contained.   For each event of this category, the primary
vertex is required to be $\geq$ 80 cm (one hadronic interaction length) from
exterior surfaces of the calorimeter, and the final state must contain
a non-scattering, exiting track with ionization compatible with a muon
mass.  This sample is useful
because the assignment of \numu\ flavor is reliable to better than 98\%, and
because the events are relatively energetic and consequently ``point well" 
to the incident $\nu$ direction.  The mean energy for neutrinos which initiate
PCEs is estimated to be 4.7 GeV, to be compared to a mean energy of
1.3 GeV for \numu\ HiRes events.  The mean angular deviation of the
reconstructed \numu\ direction versus the true direction (in Monte Carlo)
is $14^{\circ}$.  Unfortunately the number of \numu\ PCEs is low, 
less than one-third the population of our \numu\ HiRes sample.

 In order to isolate PCE two-prong and multiprong topologies, the
data events (with MC events interspersed throughout) are processed
through a software filter; those which pass are scanned.  The filter
is designed to eliminate downward-stopping muons which have endpoint
decays.   In Soudan-2, an electron from a muon decay near rangeout can give
rise to a small shower of $\leq$ 10 hits from the end of a muon
track; the topology is roughly akin to that of a neutrino-induced two-prong.
Consequently care must be taken to avoid remnant up-down asymmetry in PCEs
introduced by the filter.   The problem is avoided by requiring that there
be $\geq$ 20 hits from a PCE vertex which are additional
to the muon track.

To mitigate against cosmic-ray induced backgrounds we require that
any hit in the active shield which is coincident with a PCE, must be 
clearly associated with the exiting muon track.   Occasionally it happens that
a charged pion ejected from the cavern rock is incident upon the
calorimeter.  If the pion penetrates by more than an interaction length and
then scatters inelastically, it can mimic a $\nu$ PCE topology.   Background
events of this type are removed by requiring the net momentum of the
hadronic system of a PCE to lie within the same hemisphere which contains
the candidate muon track.

\begin{figure}[htb]
\centerline{\epsfig{file=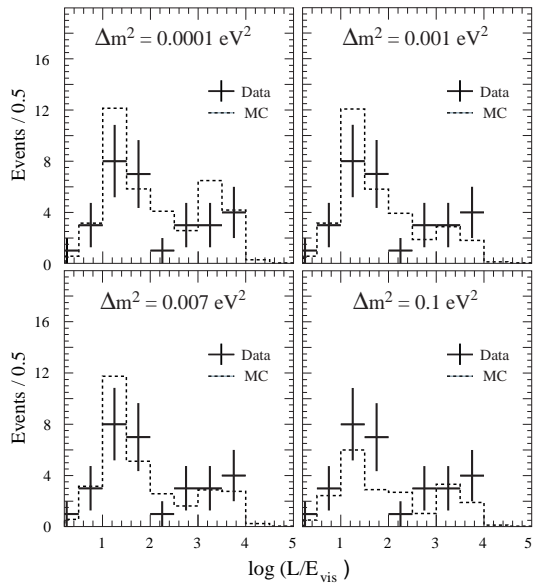,width=2.75in}}
\caption{Distribution of $\log\left(L/E_{vis}\right)$ for partially contained 
\numu\ events, compared to MC events weighted for \numu\ 
$\rightarrow$ \nutau\ oscillation, with \sqsin\ = 1 and with four different 
\delm\ values.}
\label{fig:fig07}
\end{figure}

With these selections we obtain 31 \numu\ events for which the cosmic ray
induced background is less than one event.  The \numu\ PCE rate predicted
by our Monte Carlo with no oscillations is 40 events.
The distribution of PCE data in $\log\left(L/E_{vis}\right)$ 
can be compared, as done
previously for contained HiRes \numu\ events, to representative oscillation
scenarios having \sqsin\ = 1.0.  In Fig.~\ref{fig:fig07} we observe that,
compared to the data (crosses) the oscillation prediction (dashed histogram) is
relatively high for \dlmval{10^{-4}} (Fig.~\ref{fig:fig07}a).  However the
prediction for \dlmval{10^{-1}} (Fig.~\ref{fig:fig07}d) is too low relative to
the data.  Figs.~\ref{fig:fig07}b,c suggest that the scenario preferred by the
data lies in the regime between \dlmval{1 \: \mbox{to} \: 7 \times 10^{-3}}.

\section{PLANS}

In the near future, we will include the partially contained \numu\ events into
our \sqchi\ fit to \loe .  Additionally, a sample of upward-stopping 
muon events initiated by neutrino reactions below the detector has been 
isolated and will be analyzed for oscillation effects.  And of course we will
continue to accumulate and analyze new data.  Our goal is to keep the Soudan-2
detector running and tuned for the change-of-beams incident at our underground
site.  The change will be from atmospheric $\nu$'s to Fermilab $\nu_{\mu}$'s in
Fall 2003, at which time the detector will serve the MINOS 
experiment~\cite{Wojcicki}.

\end{document}